\documentclass[10pt,a4papper]{article}
\usepackage{indentfirst}
\usepackage[]{graphicx}
\usepackage{geometry} 
\begin{document}

\title{\textbf{The decays $\tau \rightarrow (\eta,\eta')K^{-}\nu_{\tau}$ in the extended Nambu-Jona-Lasinio model}}
\author{M. K. Volkov\footnote{volkov@theor.jinr.ru}, A. A. Pivovarov\footnote{tex$\_$k@mail.ru}\\
\small
\emph{Bogoliubov Laboratory of Theoretical Physics, JINR, Dubna, 141980, Russia}}
\maketitle
\small

\begin{abstract}
The decays $\tau \rightarrow (\eta,\eta')K^{-}\nu_{\tau}$ are described in the framework of the extended Nambu-Jona-Lasinio model. Both full and differential widths of these decays are calculated. The vector and scalar channels are considered. In the vector channels, the subprocesses with intermediate $K^{*}(892)$ and $K^{*}(1410)$ mesons play the main role. In the scalar channels, the subprocesses with intermediate $K_{0}^{*}(800)$ and $K_{0}^{*}(1430)$ mesons are taken into account. Scalar channels play a less important role in calculation of the full width of the decay $\tau \rightarrow (\eta,\eta')K^{-}\nu_{\tau}$. The obtained results are in satisfactory agreement with the experimental data.
\end{abstract}
\large
\section{Introduction}
The $\tau$-decays are a good laboratory for research of strong interactions of mesons at low energies. Since the
perturbation theory of quantum chromodynamics is not applicable in this energy region (energy less than $m_{\tau} = 1.777$ GeV),
one has to use different phenomenological models. These models are generally based on the vector dominance methods
and on the chiral symmetry of the strong interactions \cite{Finkemeier:1996dh, Li:1996md, Andrianov:2005kx, Nussinov:2009sn,
Paver:2011md, Dumm:2012vb, Escribano:2013bca, Escribano:2014joa, Kang:2013jaa}.
However, most of these models include a number of fitting parameters
for a correct description of experimental data.

The Nambu-Jona-Lasinio (NJL) model \cite{Eguchi:1976iz, Ebert:1982pk, Volkov:1984kq, Volkov:1986zb, Ebert:1985kz, Vogl:1991qt, Klevansky:1992qe,
Volkov:1993jw, Ebert:1994mf, Volkov:2006vq} and its new version --- the extended NJL model \cite{Volkov:2006vq, Volkov:1996br,
Volkov:1996fk, Volkov:1997dd, Volkov:1999yi} ---  has a special place among them.
These models allow one to avoid introduction of additional arbitrary parameters. The extended NJL model is especially useful for
research of the $\tau$-lepton decays. In the case of $U(3) \times U(3)$ chiral symmetry, this model allows describing meson nonets
in both ground and first radially excited states. At the same time, due to the energy restrictions caused by the $\tau$-lepton mass,
intermediate mesons in these states play the main role in the $\tau$-lepton decays. That is why, the use of
the extended NJL model allowed a successful description of a series of $\tau$-decays, specifically
$\tau \rightarrow (\pi, \pi(1300)) \nu_{\tau}$ \cite{Ahmadov:2015zua}, $\tau \rightarrow (\eta, \eta') \pi \nu_{\tau}$
\cite{Volkov:2012be}, $\tau \rightarrow \pi \omega \nu_{\tau}$ \cite{Volkov:2012gv},
$\tau \rightarrow (\eta, \eta') 2\pi$ \cite{Volkov:2013zba}, $\tau \rightarrow (\rho(770), \rho(1450)) \nu_{\tau}$,
$\tau \rightarrow (K^{*}(892), K^{*}(1410)) \nu_{\tau}$ \cite{Ahmadov:2015oca},
$\tau \rightarrow K^{-} \pi^{0} \nu_{\tau}$ \cite{Volkov:2015vij}.

In the present work, these advantages of the extended NJL model are shown by means of the decay
$\tau \rightarrow \eta K^{-}\nu_{\tau}$. Recently, this process is actively investigated from both
experimental \cite{Inami:2008ar, delAmoSanchez:2010pc} and theoretical point of view. One of the most interesting theoretical
works on this theme is \cite{Escribano:2013bca}. In that paper, the width of the decay
$\tau \rightarrow \eta K^{-}\nu_{\tau}$ was calculated in the framework of
Chiral Perturbation Theory extended by including resonances as active fields and the prediction of the width of
$\tau \rightarrow \eta' K^{-}\nu_{\tau}$ was made. In that work, it is shown that the use of the Breit-Wigner
parametrization does not give the satisfactory results in the description of these processes. At the same time,
two other methods based on the exponential resummation and dispersive representation provide good fits.

In our work, it is shown that in the framework of the extended NJL model the use of the Breit-Wigner relation
in the standard form for description of the intermediate states leads to the satisfactory results. This statement
is confirmed by another calculations pointed above. We also make prediction of the width of decay
$\tau \rightarrow \eta' K^{-}\nu_{\tau}$, which agrees with the prediction in \cite{Escribano:2013bca}.

\section{The Lagrangian of the extended NJL model for the mesons \\
$K^{\pm}, \eta, \eta', K^{*\pm}, K_{0}^{*\pm}$ and their first radially excited states}
In the extended NJL model, the quark-meson interaction Lagrangian for pseudoscalar $K^{\pm}, \eta, \eta'$, scalar $K_{0}^{*\pm}$,
vector $K^{*\pm}$ mesons and their first radially excited states takes the form:

\begin{displaymath}
\Delta L_{int}(q,\bar{q},\eta,\eta'K,K_{0}^{*},K^{*}) = \bar{q}\left[i\gamma^{5}\sum_{j = \pm}\lambda_{j}(a_{K}K^{j} + b_{K}\hat{K}^{j})
+ \sum_{j = \pm}\lambda_{j}(a_{K_{0}^{*}}K_{0}^{*j} + b_{K_{0}^{*}}\hat{K}_{0}^{*j})\right.
\end{displaymath}
\begin{equation}
\left.+ \frac{1}{2}\gamma^{\mu}\sum_{j = \pm}\lambda_{j}(a_{K^{*}}K^{*j}_{\mu} + b_{K^{*}}\hat{K}^{*j}_{\mu})
+ i\gamma^{5} \sum_{j = u, s} \lambda_{j}
\sum_{\tilde{\eta} = \eta, \eta^{'}, \hat{\eta}, \hat{\eta}^{'}} A_{\tilde{\eta}}^{j} \tilde{\eta}\right]q,
\end{equation}
where $q$ and $\bar{q}$ are the u-, d- and s- constituent quark fields with masses $m_{u} = m_{d} = 280$MeV,
$m_{s} = 420$MeV \cite{Volkov:1999yi},\cite{Volkov:2001ns}, $\eta$, $\eta'$, $K^{\pm}$, $K^{*\pm}_{0}$ and $K^{*\pm}$ are
the pseudoscalar, scalar and vector mesons, the excited states are marked with hat,

\begin{displaymath}
a_{a} = \frac{1}{\sin(2\theta_{a}^{0})}\left[g_{a}\sin(\theta_{a} + \theta_{a}^{0}) +
g_{a}^{'}f_{a}(\vec{k}^{2})\sin(\theta_{a} - \theta_{a}^{0})\right],
\end{displaymath}
\begin{equation}
\label{Coefficients}
b_{a} = \frac{-1}{\sin(2\theta_{a}^{0})}\left[g_{a}\cos(\theta_{a} + \theta_{a}^{0}) +
g_{a}^{'}f_{a}(\vec{k}^{2})\cos(\theta_{a} - \theta_{a}^{0})\right],
\end{equation}
\begin{displaymath}
A_{\tilde{\eta}}^{j} = g_{1}^{j}b_{\tilde{\eta}1}^{j} + g_{2}^{j}f_{j}(\vec{k}^{2})b_{\tilde{\eta}2}^{j},
\end{displaymath}
$f\left(\vec{k}^{2}\right) = 1 + d \vec{k}^{2}$ is the form factor for description of the first radially excited states
\cite{Volkov:1996br},\cite{Volkov:1996fk}, $d$ is the slope parameter, $\theta_{a}$ and $\theta_{a}^{0}$ are
the mixing angles for the strange mesons in the ground and excited states

\begin{displaymath}
d_{uu} = -1.784 \textrm{GeV}^{-2}, \quad d_{us} = -1.761 \textrm{GeV}^{-2}, \quad d_{ss} = -1.737 \textrm{GeV}^{-2},
\end{displaymath}
\begin{equation}
\begin{array}{ccc}
\theta_{K} = 58.11^{\circ},     & \theta_{K_{0}^{*}} = 74^{\circ},      & \theta_{K^{*}} = 84.74^{\circ},\\
\theta_{K}^{0} = 55.52^{\circ}, & \theta_{K_{0}^{*}}^{0} = 60^{\circ},  & \theta_{K^{*}}^{0} = 59.56^{\circ}.
\end{array}
\end{equation}

The insertion of the pseudoscalar isoscalar fields requires consideration of the mixing of the four
different states: $\eta, \eta'(958), \eta(1295), \eta(1475)$, which are marked as $\eta, \eta', \hat{\eta}, \hat{\eta}'$.
The last two ones are considered as the first radially excited states of the $\eta$ and $\eta'$ mesons;
$b_{\tilde{\eta}1}^{j}$ and $b_{\tilde{\eta}2}^{j}$ are the mixing coefficients shown in Table~\ref{TabCoeff} \cite{Volkov:1999yi}.

\begin{table}[h]
\caption{The mixing coefficients for the $\eta$-mesons.}
\label{TabCoeff}
\begin{center}
\begin{tabular}{ccccc}
                        & $\eta$ & $\hat{\eta}$ & $\eta'$ & $\hat{\eta}'$ \\
$b_{\tilde{\eta}1}^{u}$ & 0.71   & 0.62         & -0.32   & 0.56          \\
$b_{\tilde{\eta}2}^{u}$ & 0.11   & -0.87        & -0.48   & -0.54         \\
$b_{\tilde{\eta}1}^{s}$ & 0.62   & 0.19         & 0.56    & -0.67         \\
$b_{\tilde{\eta}2}^{s}$ & 0.06   & -0.66        & 0.30    & 0.82
\end{tabular}
\end{center}
\end{table}

These coefficients were successfully applied for description of a series of processes with the $\eta$-mesons
\cite{Volkov:1999yi, Volkov:2013zba, Ahmadov:2013ksa}.

The matrices

\begin{displaymath}
\lambda_{+} = \frac{\lambda_{4} + i\lambda_{5}}{\sqrt{2}} = \sqrt{2} \left(\begin{array}{ccc}
0 & 0 & 1\\
0 & 0 & 0\\
0 & 0 & 0
\end{array}\right), \quad
\lambda_{-} = \frac{\lambda_{4} - i\lambda_{5}}{\sqrt{2}} = \sqrt{2} \left(\begin{array}{ccc}
0 & 0 & 0\\
0 & 0 & 0\\
1 & 0 & 0
\end{array}\right),
\end{displaymath}

\begin{equation}
\lambda_{u} = \frac{\sqrt{2}\lambda_{0} + \lambda_{8}}{\sqrt{3}} = \left(\begin{array}{ccc}
1 & 0 & 0\\
0 & 1 & 0\\
0 & 0 & 0
\end{array}\right), \quad
\lambda_{s} = \frac{-\lambda_{0} + \sqrt{2}\lambda_{8}}{\sqrt{3}} = -\sqrt{2} \left(\begin{array}{ccc}
0 & 0 & 0\\
0 & 0 & 0\\
0 & 0 & 1
\end{array}\right),
\end{equation}

\begin{displaymath}
\lambda_{0} = \sqrt{\frac{2}{3}} = \left(\begin{array}{ccc}
1 & 0 & 0\\
0 & 1 & 0\\
0 & 0 & 1
\end{array}\right),
\end{displaymath}
$\lambda_{4}$, $\lambda_{5}$ and $\lambda_{8}$ are the Gell-Mann matrices.

The coupling constants:

\begin{displaymath}
g_{K} = \left(\frac{4}{Z_{K}}I_{2}(m_{u},m_{s})\right)^{-1/2} \approx 3.77,
\quad g_{K}^{'} = \left(4I_{2}^{f_{us}^{2}}(m_{u},m_{s})\right)^{-1/2} \approx 4.69,
\end{displaymath}
\begin{displaymath}
g_{K_{0}^{*}} = \left(4I_{2}(m_{u},m_{s})\right)^{-1/2} \approx 2.78,
\quad g_{K_{0}^{*}}^{'} = \left(4I_{2}^{f_{us}^{2}}(m_{u},m_{s})\right)^{-1/2} \approx 4.69,
\end{displaymath}
\begin{displaymath}
g_{K^{*}} = \left(\frac{2}{3}I_{2}(m_{u},m_{s})\right)^{-1/2} \approx 6.81,
\quad g_{K^{*}}^{'} = \left(\frac{2}{3}I_{2}^{f_{us}^{2}}(m_{u},m_{s})\right)^{1/2} \approx 11.49,
\end{displaymath}
\begin{displaymath}
g_{1}^{u} = \left(\frac{4}{Z_{\pi}}I_{2}(m_{u},m_{u})\right)^{-1/2} \approx 3.02,
\quad g_{2}^{u} = \left(4I_{2}^{f_{uu}^{2}}(m_{u},m_{u})\right)^{-1/2} \approx 4.03,
\end{displaymath}
\begin{equation}
\label{Constants}
g_{1}^{s} = \left(\frac{4}{Z_{s}}I_{2}(m_{s},m_{s})\right)^{-1/2} \approx 4.41,
\quad g_{2}^{s} = \left(4I_{2}^{f_{ss}^{2}}(m_{s},m_{s})\right)^{-1/2} \approx 5.39,
\end{equation}
where

\begin{displaymath}
Z_{\pi} = \left(1 - 6\frac{m^{2}_{u}}{M^{2}_{a_{1}}}\right)^{-1} \approx 1.45, \quad
Z_{s} = \left(1 - 6\frac{m^{2}_{s}}{M^{2}_{f_{1}}}\right)^{-1} \approx 2.09,
\end{displaymath}
\begin{equation}
Z_{K} = \left(1 - \frac{3}{2}\frac{(m_{u} + m_{s})^{2}}{M^{2}_{K_{1}}}\right)^{-1} \approx 1.83,
\end{equation}
$Z_{\pi}$ is the factor corresponding to the $\eta - a_{1}$ transitions,
$Z_{K}$ is the factor corresponding to the $K - K_{1}$ transitions,
$Z_{s}$ is the factor corresponding to the $\eta - f_{1}$ transitions,
$M_{a_{1}} = 1230$MeV, $M_{K_{1}} = 1272$MeV, $M_{f_{1}} = 1426$MeV \cite{Agashe:2014kda} are the masses
of the axial-vector $a_{1}$, $K_{1}$ and $f_{1}$ mesons, and the integral $I_{2}$ has the following form:

\begin{equation}
I_{2}^{f^{n}}(m_{1}, m_{2}) =
-i\frac{N_{c}}{(2\pi)^{4}}\int\frac{f^{n}(\vec{k}^{2})}{(m_{1}^{2} - k^2)(m_{2}^{2} - k^2)}\theta(\Lambda_{3}^{2} - \vec{k}^2)
\mathrm{d}^{4}k,
\end{equation}
$\Lambda_{3} = 1.03$ GeV is the cut-off parameter \cite{Volkov:1986zb}.

The whole these parameters were calculated earlier and are standard for the extended NJL model.

\section{The amplitude of the decay $\tau \rightarrow \eta K^{-} \nu_{\tau}$ in the extended NJL model}
\begin{figure}[h]
\center{\includegraphics[scale = 0.7]{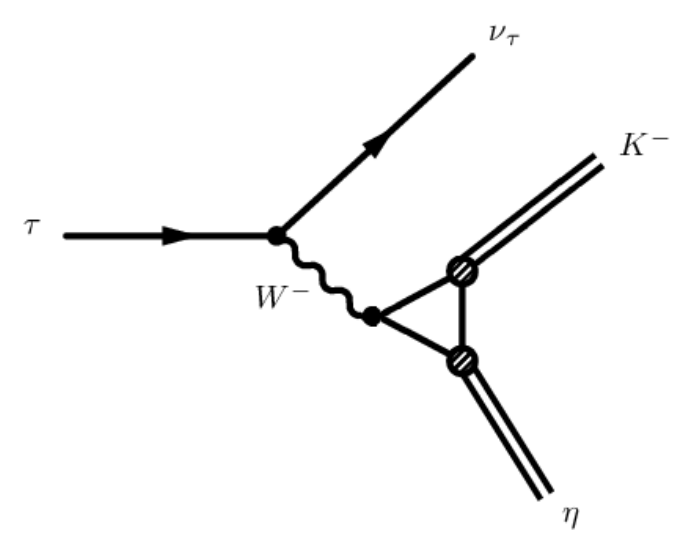}}
\caption{The decay $\tau \rightarrow \eta K^{-} \nu_{\tau}$ with the intermediate $W$-boson (Contact diagram)}
\label{Contact}
\end{figure}

\begin{figure}[h]
\center{\includegraphics[scale = 0.9]{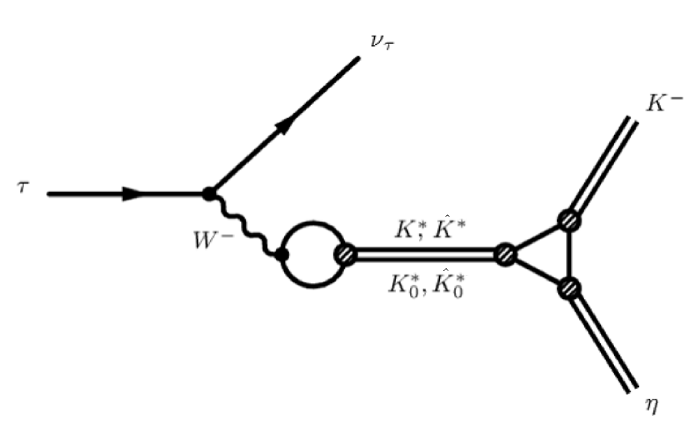}}
\caption{The decay $\tau \rightarrow \eta K^{-} \nu_{\tau}$ with the intermediate vector $K^{*}(892)$, $K^{*}(1410)$ and scalar
$K_{0}^{*}(800)$, $K_{0}^{*}(1430)$ mesons}
\label{Intermediate}
\end{figure}

The diagrams of the process $\tau \rightarrow \eta K^{-} \nu_{\tau}$ are shown in Figs.\ref{Contact},\ref{Intermediate}.


\subsection{The vector channel}
The amplitude of the process $\tau \rightarrow \eta K^{-} \nu_{\tau}$ for the vector channel takes the form:

\begin{displaymath}
T_{V} = -2i G_{F}|V_{us}|l^{\mu}
\left\{C_{1}g_{\mu\nu} + \frac{C_{2}C_{3}}{g_{K^{*}}}\cdot\frac{g_{\mu\nu}q^{2} - q_{\mu}q_{\nu}
- g_{\mu\nu}\frac{3}{2}(m_{s} - m_{u})^{2}}{M_{K^{*}}^{2} - q^{2} - i\sqrt{q^{2}}\Gamma_{K^{*}}}\right.
\end{displaymath}
\begin{equation}
\left. + \frac{C_{2}^{'}C_{3}^{'}}{g_{K^{*}}}\cdot\frac{g_{\mu\nu}q^{2} - q_{\mu}q_{\nu}
- g_{\mu\nu}\frac{3}{2}(m_{s} - m_{u})^{2}}{M_{\hat{K}^{*}}^{2} - q^{2} - i\sqrt{q^{2}}\Gamma_{\hat{K}^{*}}} \right\} (p_{K} - p_{\eta})^{\nu},
\end{equation}
where $G_{F} = 1.16637 \cdot 10^{-11}$MeV$^{-2}$ is the Fermi constant, $V_{us} = 0.2252$ is the element of the Cabbibo-Kobayashi-Maskawa matrix,
$l^{\mu} = \bar{\nu}_{\tau}\gamma^{\mu}\tau$ is the lepton current, $q = p_{K} + p_{\eta}$, $M_{K^{*}} = 896$MeV,
$M_{\hat{K}^{*}} = 1414$MeV, $\Gamma_{K^{*}} = 46$MeV, $\Gamma_{\hat{K}^{*}} = 232$MeV
are the masses and the full widths of the vector mesons \cite{Agashe:2014kda}.

The first term corresponds to the diagram with the intermediate $W$-boson, the second and the third terms correspond
to the diagrams with the intermediate vector mesons $K^{*}(892)$ and $K^{*}(1410)$. The part
$\frac{C_{2}(C_{2}^{'})}{g_{K^{*}}}\left[g_{\mu\nu}q^{2} - q_{\mu}q_{\nu} - g_{\mu\nu}\frac{3}{2}(m_{s} - m_{u})^{2}\right]$
is obtained from the quark loop in the transition of the $W$-boson into the intermediate vector meson. The part
$C_{1}(C_{3}, C_{3}^{'})(p_{K} - p_{\eta})^{\nu}$ comes from the quark triangle. The numerical coefficients

\begin{displaymath}
C_{1} = I^{a_{K}A_{\eta}^{u}} + \sqrt{2}I^{a_{K}A_{\eta}^{s}},
\end{displaymath}
\begin{displaymath}
C_{2} = \frac{1}{\sin\left(2\theta_{K^{*}}^{0}\right)}\left[\sin\left(\theta_{K^{*}} + \theta_{K^{*}}^{0}\right) +
R_{V}\sin\left(\theta_{K^{*}} - \theta_{K^{*}}^{0}\right)\right],
\end{displaymath}
\begin{displaymath}
C_{2}^{'} = \frac{-1}{\sin\left(2\theta_{K^{*}}^{0}\right)}\left[\cos\left(\theta_{K^{*}} + \theta_{K^{*}}^{0}\right) +
R_{V}\cos\left(\theta_{K^{*}} - \theta_{K^{*}}^{0}\right)\right],
\end{displaymath}
\begin{displaymath}
C_{3} = I^{a_{K}a_{K^{*}}A_{\eta}^{u}} + \sqrt{2}I^{a_{K}a_{K^{*}}A_{\eta}^{s}},
\end{displaymath}
\begin{displaymath}
C_{3}^{'} = I^{a_{K}b_{K^{*}}A_{\eta}^{u}} + \sqrt{2}I^{a_{K}b_{K^{*}}A_{\eta}^{s}},
\end{displaymath}
where
\begin{displaymath}
R_{V} = \frac{I_{2}^{f_{us}}(m_{u},m_{s})}{\sqrt{I_{2}(m_{u},m_{s})I_{2}^{f_{us}^{2}}(m_{u},m_{s})}},
\end{displaymath}
\begin{displaymath}
I^{abc} =
-i\frac{N_{c}}{(2\pi)^{4}}\int\frac{a(\vec{k}^{2})b(\vec{k}^{2})c(\vec{k}^{2})}{(m_{s}^{2} - k^2)(m_{u}^{2} - k^2)}
\theta(\Lambda_{3}^{2} - \vec{k}^2) \mathrm{d}^{4}k,
\end{displaymath}
where $a(\vec{k}^{2})$, $b(\vec{k}^{2})$ and $c(\vec{k}^{2})$ are the coefficients from the Lagrangian defined in (\ref{Coefficients}).

\subsection{The scalar channel}
The amplitude of the process $\tau \rightarrow \eta K^{-} \nu_{\tau}$ for the scalar channel takes the form:

\begin{equation}
T_{S} = -4i G_{F}|V_{us}| l^{\mu}
\left\{\frac{C_{4}C_{5}}{g_{K_{0}^{*}}}\cdot\frac{m_{s} - m_{u}}{M_{K^{*}_{0}}^{2} - q^{2} - i\sqrt{q^{2}}\Gamma_{K^{*}_{0}}}
+ \frac{C_{4}^{'}C_{5}^{'}}{g_{K_{0}^{*}}}\cdot\frac{m_{s} - m_{u}}{M_{\hat{K}^{*}_{0}}^{2} - q^{2} - i\sqrt{q^{2}}\Gamma_{\hat{K}^{*}_{0}}} \right\}q_{\mu},
\end{equation}
where $M_{K_{0}^{*}} = 682$MeV, $M_{\hat{K}_{0}^{*}} = 1425$MeV, $\Gamma_{K_{0}^{*}} = 547$MeV and $\Gamma_{\hat{K}_{0}^{*}} = 270$MeV
are the masses and full widths of the scalar mesons \cite{Agashe:2014kda}.

The part $\frac{C_{4}(C_{4}^{'})}{g_{K_{0}^{*}}}(m_{s} - m_{u})$ is obtained from the quark loop in the transition of the $W$-boson
into the intermediate scalar meson. The part $C_{5}(C_{5}^{'})q_{\mu}$ comes from the quark triangle. The numerical coefficients

\begin{displaymath}
C_{4} = \frac{1}{\sin\left(2\theta_{K^{*}_{0}}^{0}\right)}\left[\sin\left(\theta_{K^{*}_{0}} + \theta_{K^{*}_{0}}^{0}\right) +
R_{V}\sin\left(\theta_{K^{*}_{0}} - \theta_{K^{*}_{0}}^{0}\right)\right],
\end{displaymath}
\begin{displaymath}
C_{4}^{'} = \frac{-1}{\sin\left(2\theta_{K^{*}_{0}}^{0}\right)}\left[\cos\left(\theta_{K^{*}_{0}} + \theta_{K^{*}_{0}}^{0}\right) +
R_{V}\cos\left(\theta_{K^{*}_{0}} - \theta_{K^{*}_{0}}^{0}\right)\right],
\end{displaymath}
\begin{displaymath}
C_{5} = m_{s}I^{a_{K}a_{K^{*}_{0}}A_{\eta}^{u}} - \sqrt{2}m_{u}I^{a_{K}a_{K^{*}_{0}}A_{\eta}^{s}},
\end{displaymath}
\begin{displaymath}
C_{5}^{'} = m_{s}I^{a_{K}b_{K^{*}_{0}}A_{\eta}^{u}} - \sqrt{2}m_{u}I^{a_{K}b_{K^{*}_{0}}A_{\eta}^{s}}.
\end{displaymath}

\section{Numerical estimations}
The contribution of the diagrams with the vector channel to the branching
of the process $\tau \rightarrow \eta K^{-}\nu_{\tau}$ is

\begin{equation}
Br(\tau \rightarrow \eta K^{-}\nu_{\tau})_{V} = 1.46 \cdot 10^{-4}.
\end{equation}

The calculated contribution of the scalar channel is
\begin{equation}
Br(\tau \rightarrow \eta K^{-}\nu_{\tau})_{S} = 0.28 \cdot 10^{-7}.
\end{equation}

The calculated branching of the whole process is
\begin{equation}
Br(\tau \rightarrow \eta K^{-}\nu_{\tau})_{tot} = 1.45 \cdot 10^{-4}.
\end{equation}

The experimental values of this branching are
\begin{displaymath}
Br(\tau \rightarrow \eta K^{-}\nu_{\tau})_{exp} = (1.58 \pm 0.14) \cdot 10^{-4}, \textrm{ \cite{Inami:2008ar}}
\end{displaymath}
\begin{displaymath}
Br(\tau \rightarrow \eta K^{-}\nu_{\tau})_{exp} = (1.42 \pm 0.18) \cdot 10^{-4}, \textrm{ \cite{delAmoSanchez:2010pc}}
\end{displaymath}
\begin{equation}
Br(\tau \rightarrow \eta K^{-}\nu_{\tau})_{exp} = (1.52 \pm 0.08) \cdot 10^{-4}. \textrm{ \cite{Agashe:2014kda}}
\end{equation}

In conclusion, let us note that we have not taken into account the dependence of the width of $K^{*}(892)$ on
the momentum. If we assume that it grows linearly in $\sqrt{q^{2}}$, then in the considered energy region one can
put $\Gamma_{K^{*}} \approx 70$ MeV. Then the whole branching is

\begin{equation}
Br(\tau \rightarrow \eta K^{-}\nu_{\tau}) = 1.54 \cdot 10^{-4}.
\end{equation}

\begin{figure}[h]
\center{\includegraphics[scale = 0.9]{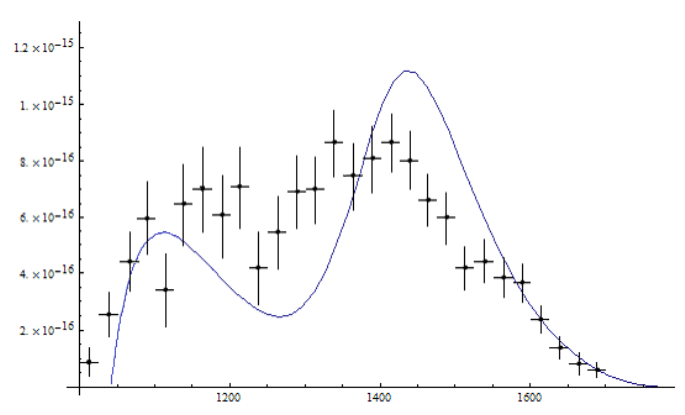}}
\caption{Differential width of the decay $\tau \rightarrow \eta K^{-} \nu_{\tau}$}
\label{Diff1}
\end{figure}

The comparison of the calculated and experimental differential width is shown in Fig.~\ref{Diff1}.
The solid line corresponds to our theoretical differential width. The points correspond to the experimental values \cite{Inami:2008ar}.

The main peak is higher than the experimental values and the theoretical line is lower than the experimental points
between two peaks. This may be due to the wrong choice of the width of $K^{*}(892)$ in the considered energy region.
The right shift of the theoretical main peak may be explained by a wrong choice of the mass of $K^{*}(1410)$.
Indeed, at the present, time there are two different experimental values for this mass: $M_{\hat{K}^{*}} = 1380 \pm 40$MeV
\cite{Aston:1987ir}, $M_{\hat{K}^{*}} = 1420 \pm 17$MeV \cite{Aston:1986jb}.

The prediction of the branching of the process $\tau \rightarrow \eta' K^{-} \nu_{\tau}$ is obtained similarly
\begin{equation}
Br(\tau \rightarrow \eta' K^{-}\nu_{\tau}) = 1.25 \cdot 10^{-6}.
\end{equation}

The experimental value is \cite{Agashe:2014kda}
\begin{equation}
Br(\tau \rightarrow \eta' K^{-}\nu_{\tau})_{exp} < 2.4 \cdot 10^{-6}.
\end{equation}

\begin{figure}[h]
\center{\includegraphics[scale = 0.6]{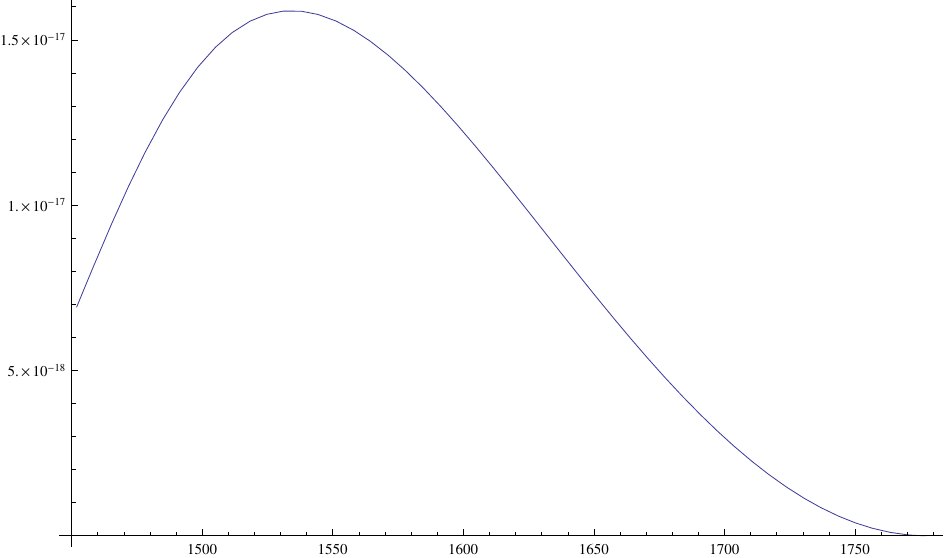}}
\caption{Differential width of the decay $\tau \rightarrow \eta' K^{-} \nu_{\tau}$}
\label{Diff2}
\end{figure}

The prediction of the differential width of the process $\tau \rightarrow \eta' K^{-} \nu_{\tau}$ is shown in Fig.\ref{Diff2}

\section{Conclusion}
The calculations carried out in the framework of the extended NJL model show that the main contribution to the width of the
decay $\tau \rightarrow \eta K^{-} \nu_{\tau}$ is given by the vector channel. The subprocesses with the
intermediate $K^{*}(892)$ and $K^{*}(1410)$ mesons play the principal role. The scalar mesons give a negligible contribution.
The prediction of the width of the process $\tau \rightarrow \eta' K^{-} \nu_{\tau}$ was made. The obtained results are in satisfactory
agreement with the experimental data.

The experimentally observed small bump of the differential width in the region of $1600$ MeV may be explained
by existing of the second radially excited state $K^{*}(1680)$, which was not taken into account here.

\section*{Acknowledgments}
We are grateful to A. B. Arbuzov and O. V. Teryaev for useful discussions.

\end{document}